\documentclass[prb,twocolumn,showpacs,amsmath,amssymb,superscriptaddress]{revtex4-1}
\usepackage{dcolumn}
\usepackage{bm,graphicx}
\usepackage{etoolbox}
\usepackage{url}
\usepackage[colorlinks]{hyperref}
\usepackage{breakurl}
\usepackage{color}
\usepackage [latin1]{inputenc}
\usepackage{appendix}

\begin{document}

\title{Landau level spectroscopy of Bi$_2$Te$_3$}

\author{I.~Mohelsk\'y}
\affiliation{Institute of Physical Engineering, Brno University of Technology, Technick\'a 2, 616 69, Brno, Czech Republic}
\affiliation{Laboratoire National des Champs Magn\'etiques Intenses, CNRS-UGA-UPS-INSA-EMFL, 25 rue des Martyrs, 38042 Grenoble, France}

\author{A.~Dubroka}
\affiliation{Department of Condensed Matter Physics and Central European Institute of Technology, Masaryk University, K\!otl\'a\v rsk\'a 2, 611 37 Brno, Czech Republic}

\author{J.~Wyzula}
\affiliation{Laboratoire National des Champs Magn\'etiques Intenses, CNRS-UGA-UPS-INSA-EMFL, 25 rue des Martyrs, 38042 Grenoble, France}

\author{A.~Slobodeniuk}
\affiliation{Laboratoire National des Champs Magn\'etiques Intenses, CNRS-UGA-UPS-INSA-EMFL, 25 rue des Martyrs, 38042 Grenoble, France}
\affiliation{Charles University, Faculty of Mathematics and Physics, Ke Karlovu 5, 121 16 Prague 2, Czech Republic}

\author{G.~Martinez}
\affiliation{Laboratoire National des Champs Magn\'etiques Intenses, CNRS-UGA-UPS-INSA-EMFL, 25 rue des Martyrs, 38042 Grenoble, France}

\author{Y.~Krupko}
\affiliation{Laboratoire National des Champs Magn\'etiques Intenses, CNRS-UGA-UPS-INSA-EMFL, 25 rue des Martyrs, 38042 Grenoble, France}
\affiliation{Institut d'Electronique et des Systemes, UMR CNRS 5214, Universit\'e de Montpellier, 34000, Montpellier, France}

\author{B.~A.~Piot}
\affiliation{Laboratoire National des Champs Magn\'etiques Intenses, CNRS-UGA-UPS-INSA-EMFL, 25 rue des Martyrs, 38042 Grenoble, France}

\author{O.~Caha}
\affiliation{Department of Condensed Matter Physics and Central European Institute of Technology, Masaryk University, K\!otl\'a\v rsk\'a 2, 611 37 Brno, Czech Republic}

\author{J.~Huml\'\i\v{c}ek}
\affiliation{Department of Condensed Matter Physics and Central European Institute of Technology, Masaryk University, K\!otl\'a\v rsk\'a 2, 611 37 Brno, Czech Republic}

\author{G.~Bauer}
\affiliation{Institut f\"ur Halbleiter- und Festk\"orperphysik, Johannes Kepler Universit\"at, Altenbergerstrasse 69, 4040 Linz, Austria}

\author{G.~Springholz}
\affiliation{Institut f\"ur Halbleiter- und Festk\"orperphysik, Johannes Kepler Universit\"at, Altenbergerstrasse 69, 4040 Linz, Austria}

\author{M.~Orlita}\email{milan.orlita@lncmi.cnrs.fr}
\affiliation{Laboratoire National des Champs Magn\'etiques Intenses, CNRS-UGA-UPS-INSA-EMFL, 25 rue des Martyrs, 38042 Grenoble, France}
\affiliation{Charles University, Faculty of Mathematics and Physics, Ke Karlovu 5, 121 16 Prague 2, Czech Republic}

\date{\today}

\begin{abstract}
Here we report on Landau level spectroscopy in magnetic fields up to 34~T performed on a thin film of the topological insulator Bi$_2$Te$_3$ epitaxially grown on a BaF$_2$ substrate.
The observed response is consistent with the picture of a direct-gap semiconductor in which charge carriers closely resemble massive Dirac particles.
 The fundamental band gap reaches $E_g=(175\pm 5)$~meV at low temperatures and it is not located on the trigonal axis, thus displaying either six or twelvefold valley degeneracy. Interestingly, our magneto-optical data do not indicate any band inversion at the direct gap. This suggests that the fundamental band gap is relatively distant from the $\Gamma$ point where profound inversion exists and
gives rise to the relativistic-like surface states of Bi$_2$Te$_3$.
\end{abstract}

\maketitle

\section{Introduction}

Bismuth telluride (Bi$_2$Te$_3$) is nowadays a widely explored material in the condensed-matter community.  Intensive investigations of Bi$_2$Te$_3$ started more than fifty years ago and they were, to a great extent,
driven by its remarkable thermoelectric properties~\cite{WrightNature58,GoldsmidM14,WittingAEM19}.
More recently, Bi$_2$Te$_3$ appeared among the very first experimentally verified
three-dimensional topological insulators which host a relativistic-type conical band on the surface~\cite{HsiehNature09,ChenScience09,HsiehPRL09,HasanRMP10}. Such surface states appear in Bi$_2$Te$_3$
due to band inversion at the center of the Brillouin zone where spin-orbit interaction reverses the
order of $p$-states as compared to isolated atoms of tellurium and bismuth~\cite{ZhangNaturePhys09}.

Despite considerable experimental effort, the electronic band structure of Bi$_2$Te$_3$ is nowadays only partly understood.
The current consensus implies that Bi$_2$Te$_3$ is a narrow-gap semiconductor. Nevertheless,
the number and positions of extrema in the lowest-lying conduction and top-most valence bands, still remain under debate. Hence, it is still not clear whether the energy band gap is direct -- with the extrema in the conduction
and valence band aligned in the momentum space -- or indirect.

\begin{figure}[t]
      \includegraphics[width=.45\textwidth]{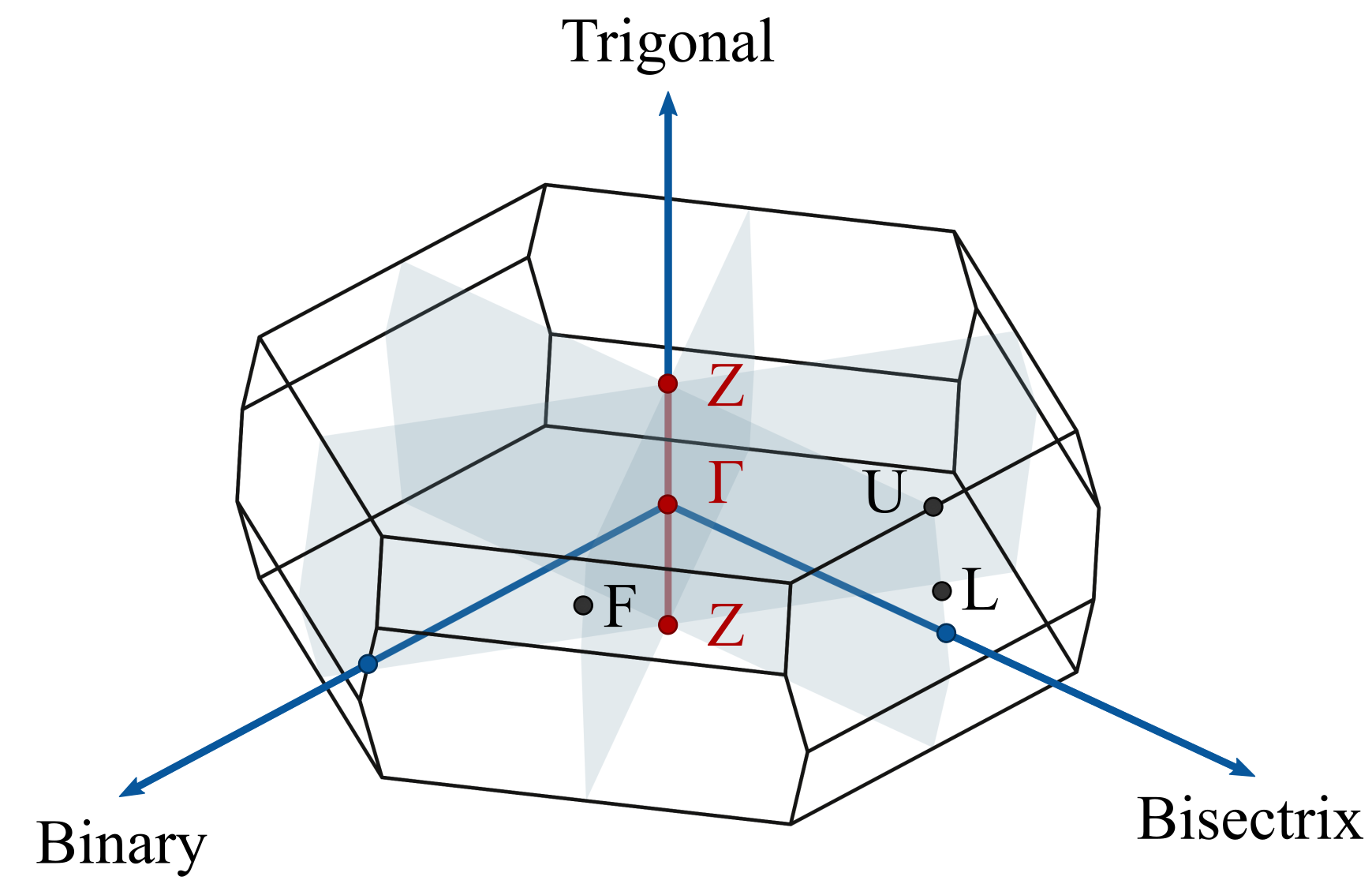}
      \caption{\label{BZ} Schematic view of the first Brillouin zone of Bi$_2$Te$_3$ with the mirror planes (gray planes) and trigonal, bisectrix as well as binary axes indicated.}
\end{figure}

From quantum oscillation experiments~\cite{DrabblePPS56,MallinsonPR68,Kohlerpssb76,Kohlerpssb76II,Kohlerpssb76III, KulbachinskiiPRB94,RischauPRB13}, it was concluded that six non-equivalent valleys exist both in the conduction as well as in the valence band, located pairwise in the mirror planes of Bi$_2$Te$_3$ (cf. Fig.~\ref{BZ}).
ARPES studies~\cite{ChenScience09,AlpichshevPRL10,ChenPNAS12,Sanchez-BarrigaPRB14,ChuangRCSA18}
also suggest multiple, likely sixfold~\cite{LiAM10}, valley degeneracy in the valence band, but the minimum of the  conduction band appears to be projected to
the $\overline{\Gamma}$ point of the surface Brillouin zone. This implies either none or double valley degeneracy, with the minimum at the $\Gamma$ ($Z$) point or on the $\Gamma-Z$ line, respectively. In optical experiments~\cite{AustinPPS58,SehrJPCS62,GreenawayJPCS65,ThomasPRB92,VilaplanaPRB11,ChaplerPRB14,DubrokaPRB17,PeirisJVST19}, both direct and indirect band gaps were reported with widths not exceeding $200$~meV. Available theoretical studies provide us with diverse views on the electronic bands in Bi$_2$Te$_3$~\cite{MishraJPCM97,YounPRB01,LarsonPRB03,WangPRB07,ZhangNaturePhys09}. Presumably more accurate GW calculations predict multiple extrema of the highest valence band and one or more valleys in the conduction band~\cite{YazyevPRB12,AguileraPRB13,NechaevPRB13,MichiardiPRB14,AguileraPRB19}. The calculated band gap thus can be both direct or indirect. Its magnitude depends on the used functional and falls into a relatively broad range of energies between 50 and 200~meV.

In this paper, we study the bulk magneto-optical response of a thin layer of Bi$_2$Te$_3$.
We show that the relatively complex response, comprising a series
of interband and intraband inter-Landau level (inter-LL) excitations,  may be explained using a simple two-band model for a time-reversal-invariant direct-gap semiconductor. This implies that the charge carriers in Bi$_2$Te$_3$ behave, to a certain extent,  as massive Dirac electrons.
The selection rules for electric-dipole excitations observed experimentally allow us to deduce the symmetry of electronic bands around the fundamental band gap. In this way, we conclude that the direct band gap is located away from the trigonal axis,
and therefore, it displays a multiple degeneracy ($N=6$ or 12).

\section{Sample preparation and experimental details}

The studied Bi$_2$Te$_3$ epilayer with a thickness of $300$~nm was grown using molecular beam epitaxy on a 1-mm-thick (111)-oriented cleaved BaF$_2$ substrate~\cite{CahaCGD13,SteinerJAC14}. The details about the growth technique and the sample characterization were presented previously~\cite{DubrokaPRB17}. The Hall measurements at liquid-helium temperature, performed on the same sample as all our optical and magneto-optical studies, indicate $p$-type conductivity and a moderate hole density of $p\sim 2\times 10^{18}$~cm$^{-3}$.

To measure infrared magneto-transmission, nonpolarized radiation from a globar or a mercury lamp was analyzed by a commercial Bruker Vertex 80v Fourier-transform spectrometer. The radiation was then delivered via light-pipe optics to the sample kept in the helium exchange gas at the temperature $T=2$~K and placed in a superconducting solenoid or in the resistive high-field magnet (above 13 T, up to 34 T), both at the LNCMI in Grenoble.
The light transmitted through the sample was detected by a composite bolometer, placed directly below the sample.
The studied sample was probed in the Faraday configuration, with the magnetic field applied along the trigonal axis (rhombohedral or $c$-axis) of Bi$_2$Te$_3$.
The measured transmission spectra, $T_B$, were normalized by the zero-field transmission, $T_0$, and plotted in the form of relative magneto-transmission, $T_B/T_0$, or relative magneto-absorbance, $A_B=-\ln[T_B/T_0]$. The optical
response at $B=0$ was deduced from the ellipsometric measurements realized using a commercial Woollam IR-VASE ellipsometer coupled to a closed He-cycle cryostat, for details see Ref.~\onlinecite{DubrokaPRB17}.

\section{Two-band model of a direct-gap semiconductor}

To interpret our experimental data presented and discussed below, we adopt a simple two-band model for a time-reversal-invariant direct-gap semiconductor. The corresponding Hamiltonian may be derived, \emph{e.g.}, using the first-order {$\mathbf{k}\cdot\mathbf{p}$} theory applied at a particular point, $\mathbf{k_0}$, of the Brillouin zone:
\begin{equation}\label{H0}
h = \left[ \begin{matrix} \Delta  &\hbar v_D (q_x+iq_y) \\ \hbar v_D (q_x-iq_y)&-\Delta \end{matrix} \right ],
\end{equation}
where $\mathbf{q} = \mathbf{k}-\mathbf{k_0}= (q_x,q_y,0)$. The Hamiltonian describing electrons and holes with an opposite spin projection in the doubly degenerate bands is obtained by the complex conjugation of $h$ ($h^*$). Notably, we consider only vanishing $q_z$ momenta because the  $q_z=0$ states  dominate the overall optical response when the magnetic field is applied along the $z$ axis.

\begin{figure}[t]
      \includegraphics[width=.42\textwidth]{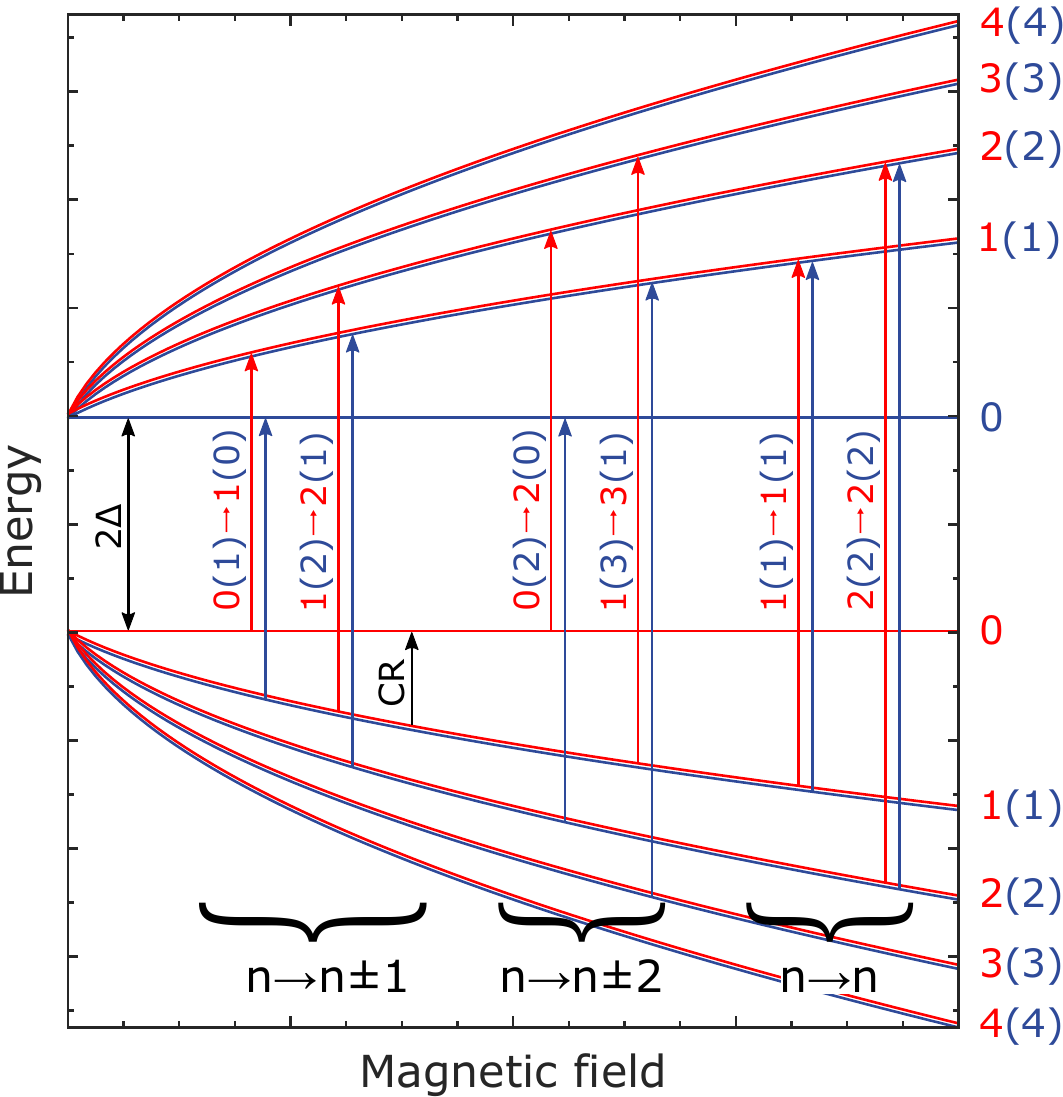}
      \caption{\label{Schematics} Schematic view of the LL spectrum corresponding to the $h$ and $h^*$ Hamiltonians (blue and red, respectively). In the two-band model, the zero-mode ($n=0$) LLs remain independent of $B$ unless the Hamiltonians are expanded to include band inversion or
      the impact of higher/lower lying bands.
      The vertical arrows correspond to electric-dipole active transitions from three distinct sets. In the system with a full rotational symmetry around the $c$ axis,  only transitions $n\rightarrow n\pm 1$ may appear. CR denotes cyclotron resonance in the quantum limit of a $p$-type system. The $n\rightarrow n\pm 2$ excitations become active when the rotational symmetry is reduced to the trigonal one. When the symmetry is further reduced, transitions $n\rightarrow n$ may emerge.}
\end{figure}

The above Hamiltonian gives rise to the conduction and valence bands with a characteristic relativistic-like hyperbolic profile: $E(k)=\pm\sqrt{\Delta^2+\hbar^2v_D^2k^2}$. They are separated by the band gap of $2\Delta$ and display the full particle-hole symmetry.
Interestingly, there exists a non-trivial analogy between the proposed two-band model and truly relativistic systems of massive electrons described by the Dirac equation~\cite{Kacmanpssb71,ZawadzkiHMF97,GoerbigEPL14,OrlitaPRL15,ZawadzkiJPCM17}. This analogy implies that the band-edge effective masses of electrons and holes are equal to the quantity referred to as the Dirac mass, $m_D=m_e=m_h=\Delta/v_D^2$. The effective (cyclotron) mass of charge carriers increases linearly with the energy distance $\varepsilon$ from the band-edge:
$m_{e,h}(\varepsilon)=m_D(1+\varepsilon/\Delta)=m_D |E|/\Delta$.
The corresponding $g$ factors are expressed as $g_e=g_h=2m_0/m_D$, where $m_0$ stands for the bare electron mass. For large momenta, the dispersion of charge carriers approaches the ultra-relativistic limit, $E(\mathbf{k})\approx \pm\hbar v_D |\mathbf{k}|$. This allows us to treat the $v_D$ parameter which describes the coupling between bands, as the effective velocity of light in the explored system.

\begin{figure}
      \includegraphics[width=.50\textwidth]{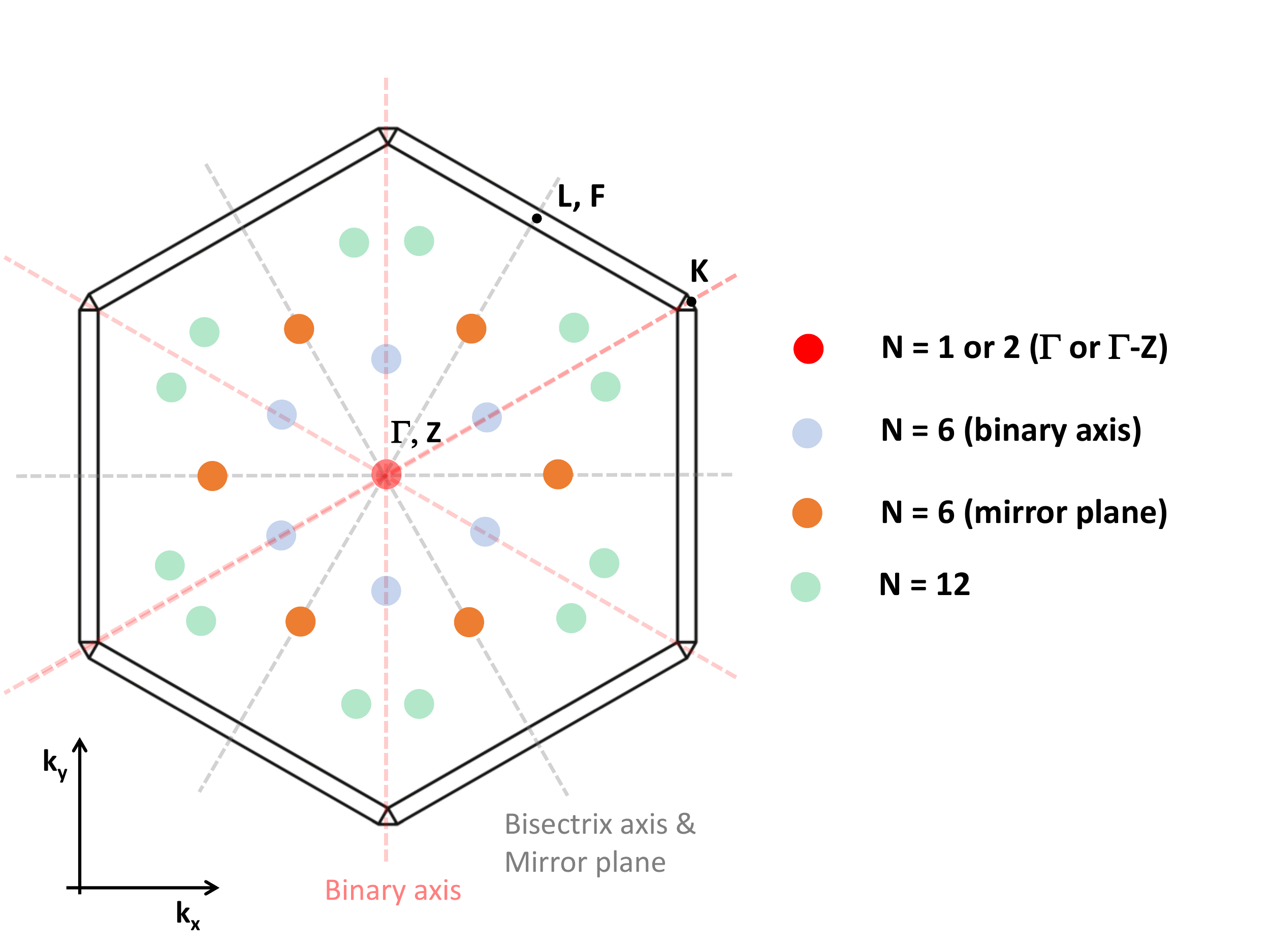}
      \caption{\label{Projected-BZ}
      The Brillouin zone of Bi$_2$Te$_3$ projected to the plane perpendicular to the trigonal axis. The full circles show possible positions of the band extrema, and their corresponding degeneracy, given by the crystal symmetry: at the $\Gamma$ point or on the $\Gamma-Z$ line ($N=1$ or 2), in the mirror planes ($N=6$), on the binary axes ($N=6$) or in a general position ($N=12$).}
\end{figure}

\begin{figure}
      \includegraphics[width=.46\textwidth]{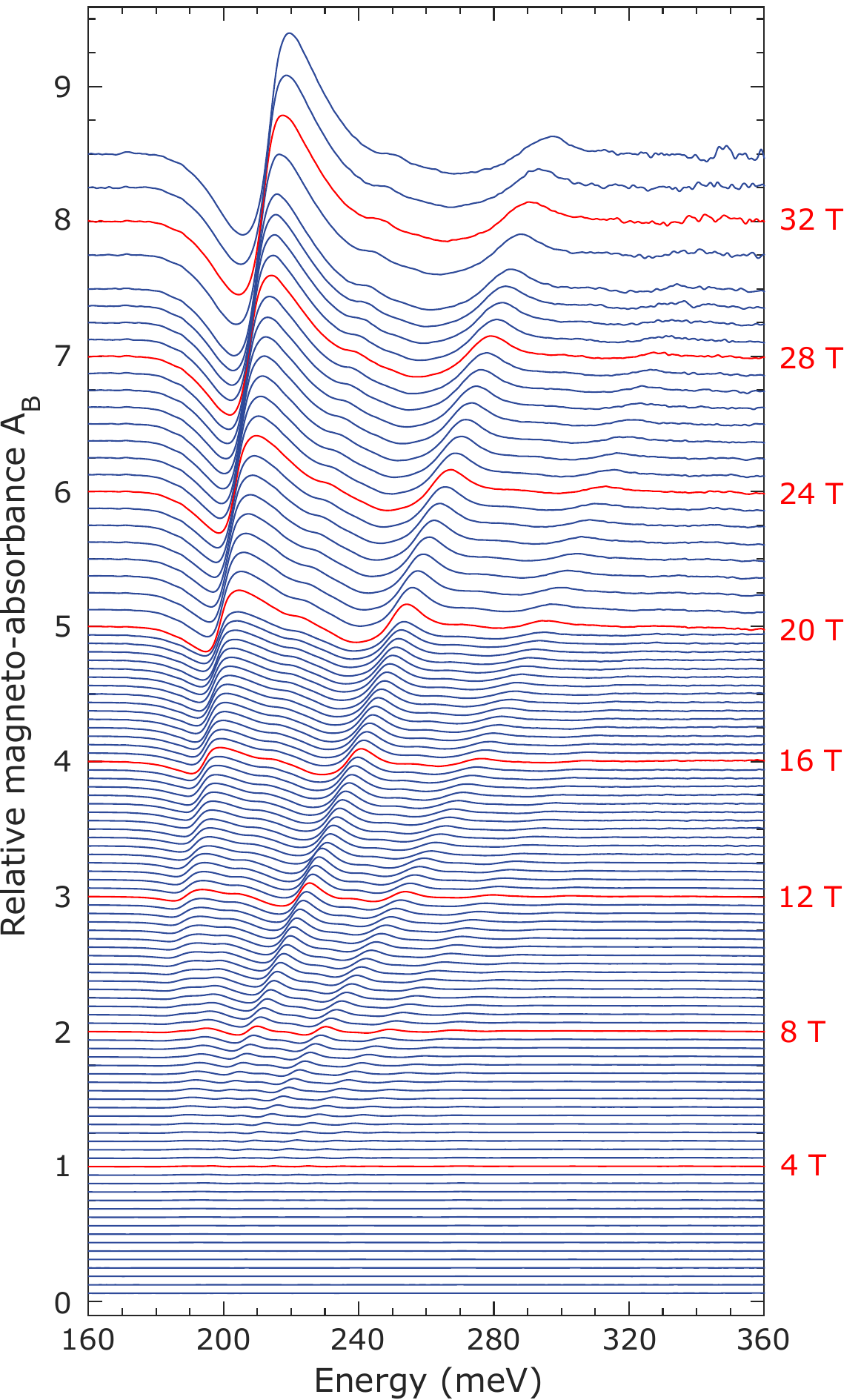}
      \caption{\label{MIR-water-fall} Relative magneto-absorbance spectra, $A_B=-\ln[T_B/T_0]$, plotted for selected values of the applied magnetic
      field. The maxima correspond to individual inter-LL resonances. The minima (negative values of $A_B$) emerge due to suppression of zero-field absorption.
       A baseline-line correction has been performed (subtraction of a linear background) to ensure $T_B/T_0 \approx 1$ in the spectral range where no inter-LL excitations were observed (at high and low photon energies). Afterwards, the baseline of each spectrum has been shifted vertically by the offset of 0.25$\times B$[T].}
\end{figure}

\begin{figure}
      \includegraphics[width=.48\textwidth]{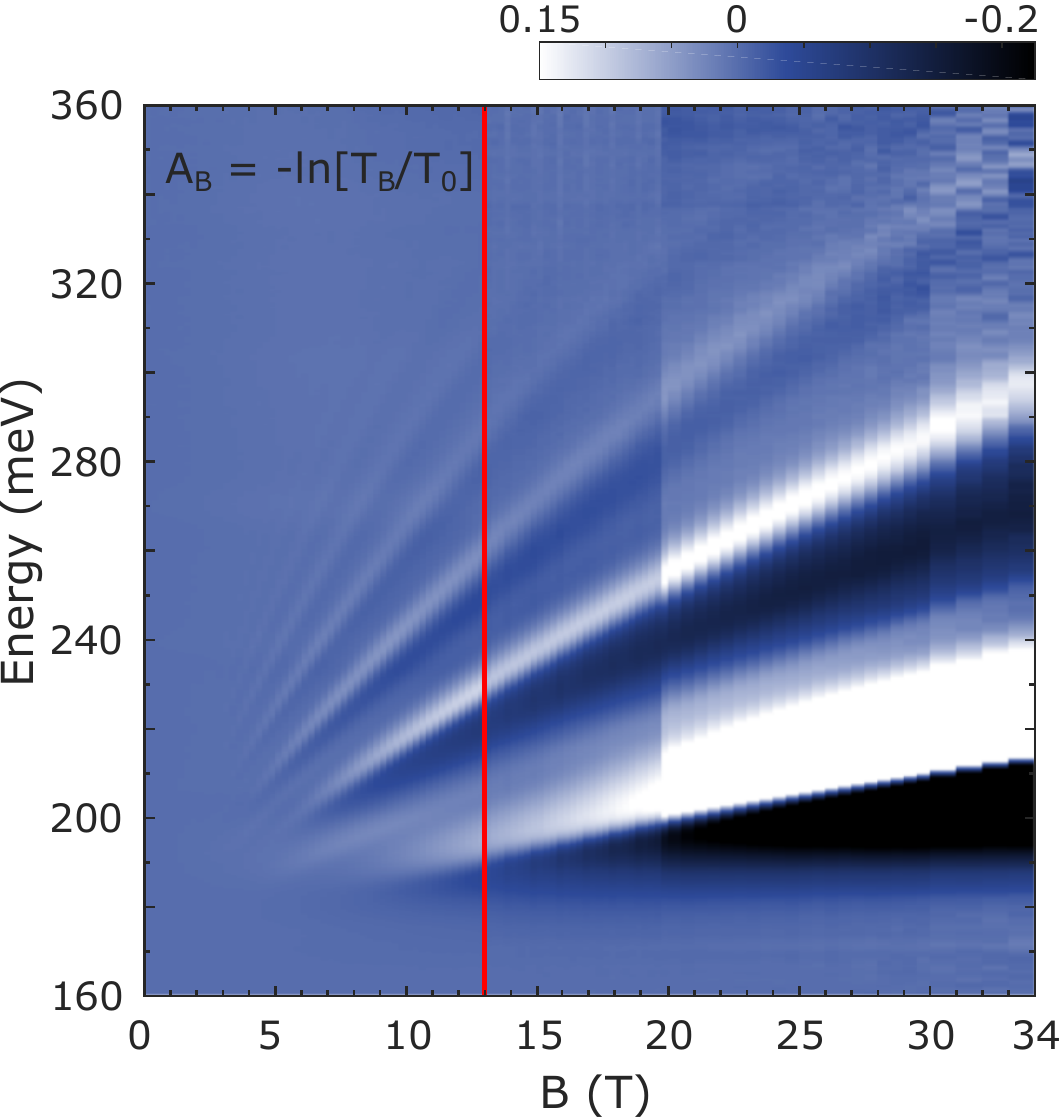}
      \caption{\label{MIR-color-plot} Relative magneto-absorbance, $A_B$, from Fig.~\ref{MIR-water-fall}, visualized as a false-color plot. The vertical red line separates the data collected using the superconducting and resistive coils, below and above 13~T, respectively.}
\end{figure}

When a magnetic field is applied, the electronic band structure transforms into Landau levels with the spectrum: $E_n=\pm\sqrt{v_D^2 2e\hbar B n + \Delta^2}$, where $n>0$. In addition, there exists a pair of spin-polarized zero-mode LLs ($n=0$) with the energies $E_0=\Delta$ and $E_0=-\Delta$,
which correspond to the $h$ and $h^*$ Hamiltonians, respectively. Our model, characterized by the full rotational symmetry around the $z$-axis, implies the standard selection rules, $n \rightarrow n\pm 1$, for electric-dipole excitations in the Faraday configuration (Fig.~\ref{Schematics}). A closer analysis~\cite{OrlitaPRL15,LyJPCM16} shows that the excitations
within the LL spectra of the $h$ and $h^*$ Hamiltonians
dominantly follow the selections rules $n \rightarrow n - 1$ and $n \rightarrow n + 1$ and they are
active in $\sigma^-$ and $\sigma^+$ polarized light, respectively. This remains valid until the energy of excitations exceeds, by an order of magnitude, the band gap of $2\Delta$.

Let us now reconcile the proposed toy model of massive Dirac electrons with the real band structure of Bi$_2$Te$_3$. We align the $z$ axis in the model with the trigonal axis of Bi$_2$Te$_3$. The bands considered in the model are associated with the lowest lying conduction and the topmost valence bands which are primarily formed from $p$-like states of bismuth ($6p$) and tellurium ($5p$)~\cite{ZhangNaturePhys09,LiuPRB10}. The electronic bands at higher or lower energies are completely neglected.
Even though we deal with a material proven to be a topological insulator~\cite{HsiehNature09,ChenScience09,HsiehPRL09}, the band inversion, putting tellurium states above bismuth ones, does not have to be present around an arbitrarily chosen point $\mathbf{k_0}$ in the Brillouin zone, especially for the momenta far from the $\Gamma$ point.

The position of the $\mathbf{k_0}$ point in the Brillouin zone (see Fig.~\ref{Projected-BZ}) and the crystal symmetry imply the valley degeneracy $N$. When the possibility of an accidental degeneracy is neglected, the valley multiplicity reaches $N=1$ when the $\mathbf{k_0}$ point coincides with the $\Gamma$ or $Z$ points (a half of the valley at both zone-boundary $Z$ points), $N=2$ for  $\mathbf{k_0}$ located on the trigonal axis (between $Z$ and $\Gamma$ points), $N=6$ for the valleys located in the mirror planes or on the binary axis (perpendicular to the trigonal axis) and $N=12$ for a general position in the Brillouin zone~\cite{DrabblePPS56}.

Importantly, the symmetry of the crystal and the position of the $\mathbf{k_0}$ point in the Brillouin zone (see Fig.~\ref{BZ}) may have a profound impact on the related magneto-optical response. The particular location of the $\mathbf{k_0}$ point in the Brillouin zone implies a specific form of high-order momentum terms that result from the {$\mathbf{k}\cdot\mathbf{p}$} expansion. Often, these terms do not impact the LL spectrum significantly, and therefore, they are not included in the Hamiltonian~\ref{H0}. Nevertheless, the lowered rotational symmetry gives rise, in principle, to additional sets of electric-dipole transitions.

For the point $\mathbf{k_0}$ located on the trigonal axis, the angular momentum of an absorbed photon is conserved only modulo 3~\cite{LyJPCM16}. Therefore, additional sets of electric-dipole transitions may emerge in the excitations spectrum: $n \rightarrow n \pm 2, 5, 7\ldots$. The situation is similar to $K$ point electrons in bulk graphite~\cite{OrlitaPRL12} for which the pronounced trigonal warping does not
alter significantly the LL spectrum, but it gives rise to a series of cyclotron resonance (CR) harmonics that would be strictly forbidden in the electric-dipole approximation in a system with full rotational symmetry.
Further sets of inter-LL excitations may appear for the $\mathbf{k_0}$ point located away from the trigonal axis of Bi$_2$Te$_3$ -- for instance, interband transitions that conserve the LL index: $n \rightarrow n$.

\section{Experimental data and discussion}

The magneto-optical data collected on the studied Bi$_2$Te$_3$ epilayer
at $T=2$~K in the mid-infrared spectral range are presented in Figs.~\ref{MIR-water-fall} and \ref{MIR-color-plot}, as a stacked-plot
and false-color plot of relative magneto-absorbance spectra, $A_B$, respectively. The data comprise a series of pronounced resonances that follow a sub-linear in $B$ dependence, and in the limit of the vanishing magnetic field, extrapolate to a finite (positive) energy. We interpret these resonances as inter-LL excitations that promote electrons across the band gap. The resonances have pronounced high-energy tails that are typical of interband inter-LL excitations in bulk systems.

\begin{figure*}[t]
      \includegraphics[width=.95\textwidth]{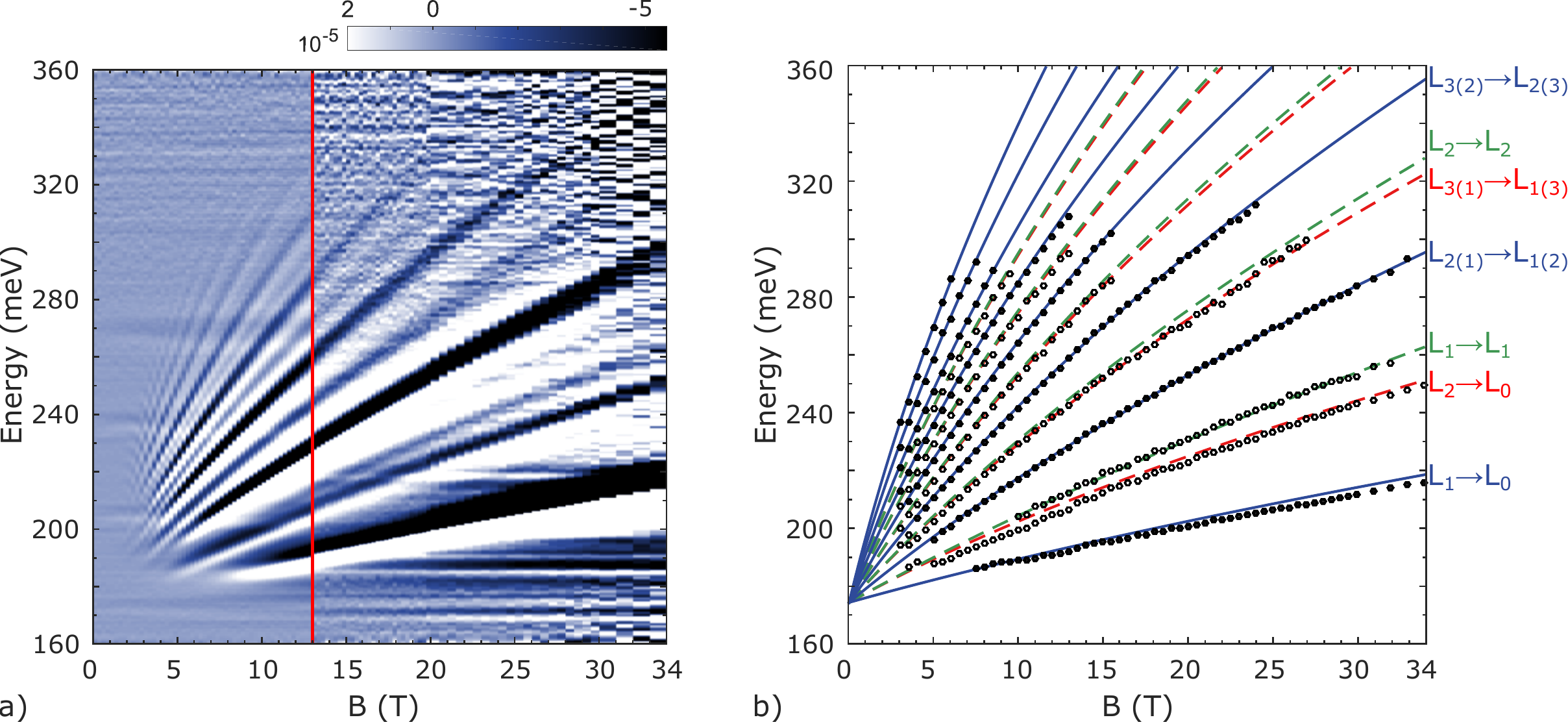}
      \caption{\label{MIR_der}
      Part (a): The false-color plot of the second derivative of relative magneto-absorbance in the middle infrared spectral range, $\mathrm{d}^2 A_B/\mathrm{d}\omega^2$. The vertical red line separates the data collected using the superconducting and resistive coils, below and above 13~T, respectively. Part (b): The deduced minima of $\mathrm{d}^2 A_B/\mathrm{d}\omega^2$ associated with the positions of interband inter-LL excitations. The solid (open) circles correspond to positions of dominant (weak) lines in the spectra. The lines correspond to the theoretically expected positions of resonances that follow three different selection rules: $n \rightarrow n\pm 1$ (solid blue), $n \rightarrow n\pm 0$ (dashed green) and $n \rightarrow n\pm 2$ (dashed red).}
\end{figure*}

A more detailed analysis shows the presence of additional transitions with a considerably weaker intensity. These latter transitions become clearly visible in the second derivative of the relative magneto-absorbance, plotted as a false-color plot (Fig.~\ref{MIR_der}a). Since the observed resonances are relatively sharp, we associate the minima in $\mathrm{d}^2 A_B/\mathrm{d}\omega^2$ curves directly with the positions of excitations (Fig.~\ref{MIR_der}b).
Using the method of second derivative, we also reduce the impact of the normalization by relatively flat zero-field transmission $T_0$ on the deduced energies of excitations. No resonances attributable to surface states were identified, in contrast to preceding studies~\cite{WolosPRL12}.

Let us now compare our experimental data with expectations based on the proposed two-band model. In the first step, we assign the series of dominant transitions,
which contains up to eight well-resolved lines, with the position of
$n \rightarrow n\pm 1$ resonances expected in a  system with the full rotational symmetry.
Very good agreement is found for an energy band gap $E_g=2\Delta=(175\pm 5)$~meV and a velocity parameter $v_D = (4.7\pm 0.1)\times 10^5$~m/s, as shown by solid lines in Fig.~\ref{MIR_der}b. In this way, all dominant lines may be explained using
only one widely tunable parameter $v_D$. This is because the band gap $2\Delta$ falls into a fairly narrow interval given by the zero-field extrapolation of lines.

\begin{figure*}[t]
      \includegraphics[width=.89\textwidth]{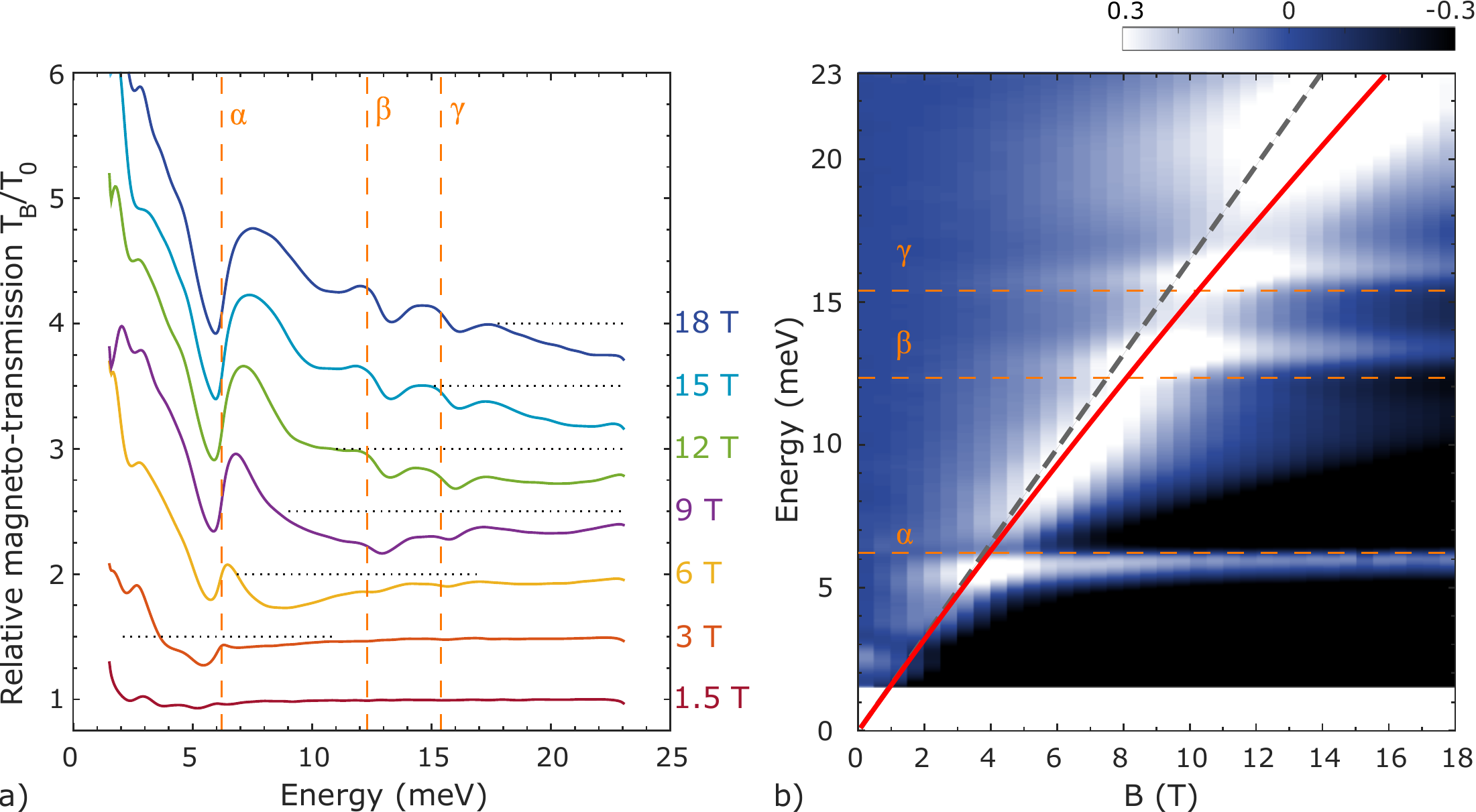}
      \caption{\label{FIR} Part (a): Relative magneto-transmission spectra, $T_B/T_0$ for plotted for selected values of $B$. The CR resonance absorption is manifested by a relatively broad minimum that shifts almost linearly with $B$ towards higher energies.
      The pronounced $B$-induced transmission at low photon energies ($T_B/T_0>1$) is typical of the CR response in the quasi-classical regime~(see, e.g.~\cite{WitowskiPRB10}). This is due to Drude-type absorption which dominates the response at $B=0$ and which is suppressed by the applied magnetic field. The horizontal dotted lines show $T_B/T_0=1$ level for each stacked spectrum.
      Part (b): False-color plot of relative magneto-absorbance $A_B$ in the far infrared range. The solid red line corresponds to the energy of the fundamental CR mode, $1\rightarrow0$, calculated within the two-band model (cf. Fig.~\ref{Schematics}). The dashed gray line shows the cyclotron energy, $\hbar\omega_c = \hbar eB/m_D$, for the band-edge (Dirac) mass $m=\Delta/v_D^2=0.07m_0$. The vertical and horizontal dashed lines in parts (a) and (b), respectively, show positions of three phonon modes ($\alpha$, $\beta$ and $\gamma$) observed in zero-field transmission spectrum. One may associate them with $E_u^1$, $E_u^2$ and $A_{1u}$ infrared-active phonons~\cite{Richterpssb77}. The possible coupling between these phonons and CR mode is discussed in the main text. The last mode is supposed to be active for the electric-field component along the $c$ axis, and it appears since the radiation is focused on the sample using a cone.}
\end{figure*}

The parameters $\Delta$ and $v_D$ deduced from the fit of dominant interband transitions allows us to predict the position and $B$-dependence of the fundamental CR mode, $1 \rightarrow 0$, in our $p$-type sample (cf. Fig.~\ref{Schematics}). This may be compared to the experimentally observed CR mode visible as a broad minimum in $T_B/T_0$ curves or the maximum in the false-color plot of relative magneto-absorbance $A_B$ (Figs.~\ref{FIR}a and b, respectively).
Moreover, the expected position (red solid line in Fig.~\ref{FIR}b) matches very well the experimental data without any additional adjustment of $v_D$ or $\Delta$.
The cyclotron energy deviates only weakly from a linear dependence in $B$. This allows us to describe it approximately by the formula
for the cyclotron energy of a parabolically dispersing particle: $\hbar\omega_c = \hbar eB /m_D$ using the Dirac mass of $m_D=0.07m_0$
(dashed gray line in Fig.~\ref{FIR}b).
This value is in very good agreement with preceding studies which reported the band-edge mass of $0.08m_0$~\cite{Kohlerpssb76II,KulbachinskiiJPSJ99}. Let us note that the observed CR mode indicates a coupling with optical phonons that may resemble the magneto-polaron effect~\cite{PeetersPRB85,PeetersPRB86}. Nevertheless, a relatively broad CR line, as compared to much sharper phonon resonances, does not allow us to analyze this effect in detail.

Importantly, our model does not include any electron-hole asymmetry. The experimentally deduced Dirac mass $m_D$ thus represents a good estimate of the band-edge mass for both, holes and electrons. Indeed, the extracted Dirac mass of $m_D=0.07m_0$ is close to the electron band-edge mass of $0.06m_0$ deduced in quantum oscillation experiments in the past~\cite{Kohlerpssb76}. Importantly, this latter agreement indicates that the conduction-band minimum, hosting the final states of the observed interband inter-LL excitations, is not just a local extremum, but the global one. The parameter $2\Delta$ thus corresponds to the fundamental band gap in Bi$_2$Te$_3$.
Hypothetically, one can imagine that there exist other extrema of the conduction and valence bands, which are characterized by the band-edge masses identical to the ones of valleys probed in our magneto-optical experiments. In such a case, the real energy band gap might still be smaller and indirect. However, we do not find such a coincidence probable.

The very good agreement between experimental data and our simple two-band model suggests that Bi$_2$Te$_3$ is a direct-gap semiconductor. However, one can never exclude a small displacement $\delta k_{c-v}$ between the extrema of the conduction and valence bands. Thanks to our low-field magneto-optical data, we may find the upper limit of such a displacement. It is approximately given by the reciprocal value of the magnetic length $\delta k_{c-v} \sim 1/l_B=\sqrt{eB/\hbar}\approx 0.07$~nm$^{-1}$, taken at the onset of LL quantization in our sample ($B\approx3$~T).
Such a distance represents only a small fraction of the whole Brillouin zone size. This allows us to speak about the direct character of the band gap in Bi$_2$Te$_3$ with reasonable justification.

The validity of the proposed model also implies that the Zeeman splitting in Bi$_2$Te$_3$ should be equal to cyclotron energy, $E_Z=E_C$, that is typical of massive Dirac electrons~\cite{Landau77,IzakiPRL19} and the $g$ factors of electrons and holes should reach $g_e=g_h=2m_0/m_D\approx 30$. In fact, large values of $g$ factors are expected in systems composed of heavy elements with strong spin-orbit coupling~\cite{CohenPM16,WolosAIP13}. To the best of our knowledge, no results from spin-resonance experiments on Bi$_2$Te$_3$ have been reported so far. Relatively large values for $g$ factors were estimated from quantum oscillation experiments~\cite{DrathPRA67,Kohlerpssb76,Kohlerpssb76III,RischauPRB13}, but the concluded ratio was somewhat lower than unity: $E_Z/E_C\approx 0.5-0.7$. The difference may be attributed to the influence of more distant bands~\cite{RothPR59}.

In a second step, we compare the positions of additional lines -- with weaker integral intensities but still clearly manifested in the spectra -- with the energies of other possibly existing inter-LL excitations. As a matter of fact, all additionally observed lines fit very well to two series: $n \rightarrow n \pm 2$ and $n \rightarrow n$ marked by dashed lines in Fig.~\ref{MIR_der}b. These two series nearly overlap for higher indices $ n $, nevertheless, at low photon energies, one may clearly distinguish two separate $2 \rightarrow 0$ and $1 \rightarrow 1$ lines. Let us notice that the $0 \rightarrow 2$ transition should not appear due to the occupation effect in our $p$-type sample.
The proposed  two-band model, implying not more than two tunable parameters, is
thus capable of explaining all intraband and interband inter-LL excitations
resolved in our magneto-optical data (more than 10 lines). This strongly corroborates our interpretation.

The additionally appearing inter-LL excitations reflect the symmetry of electronic bands at the fundamental band gap. The presence of the $n \rightarrow n$ series indicates that the symmetry is definitely lower than the trigonal one. The direct fundamental gap, evidenced in our magneto-optical experiment,
thus cannot be located on the trigonal axis of Bi$_2$Te$_3$ and the valley degeneracy must reach either $N=6$ or $N=12$, unless some accidental degeneracy occurs. Both possibilities have been discussed in the literature. The former one, corresponding to the energy gap within the mirror planes of Bi$_2$Te$_3$
(see Fig.~\ref{BZ}), has been concluded as more probable based on quantum oscillations experiments~\cite{DrabblePPS58,MallinsonPR68,Kohlerpssb76,Kohlerpssb76II,Kohlerpssb76III}, which followed the response for various orientations of the applied magnetic field with respect to crystallographic axes

\begin{figure}
      \includegraphics[width=.47\textwidth]{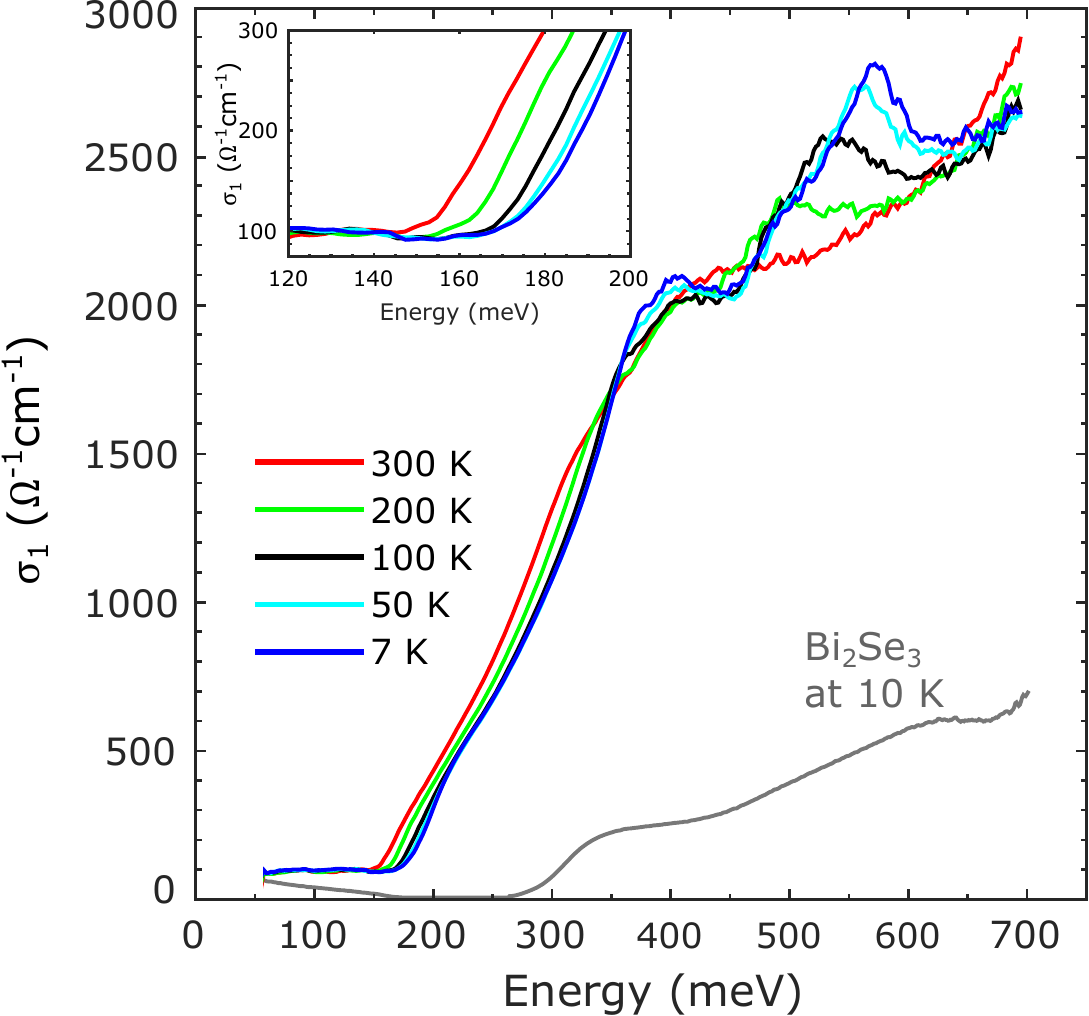}
      \caption{\label{Zero-field} The real part of optical conductivity, $\sigma_1(\omega)$, of the studied Bi$_2$Te$_3$ sample at the selected temperatures measured using the ellipsometry technique.  The inset shows in detail the sharp increase of $\sigma_1(\omega)$ due to the onset of interband excitations in Bi$_2$Te$_3$ at selected temperatures. For comparison, the optical conductivity spectrum of Bi$_2$Se$_3$ has been reprinted from~\cite{DubrokaPRB17}, see the main text.
      Due to the Moss-Burstein effect~\cite{BursteinPR54,MartinezSR17},
      the onset of interband absorption
      is around 300~meV in this particular Bi$_2$Se$_3$ sample, which is
      well above the energy band gap of 220~meV~\cite{MartinezSR17}.
      }
\end{figure}

Let us confront our conclusions -- about the size, nature and multiplicity of the fundamental band gap in Bi$_2$Te$_3$ -- with the zero-field optical response obtained using ellipsometry. In line with expectations, the deduced optical conductivity (Fig.~\ref{Zero-field}) shows a rather steep increase at photon energies slightly above the band gap $E_g=175$~meV estimated from our magneto-optical experiments. The optical response indicated the presence of several critical points in the explored part of the infrared spectral range, which are discussed in the appendix.
The lowest one, located around $E_g^*\sim 188$~meV, corresponds to the onset of interband absorption, the so-called optical band gap. The difference $E_g^*-E_g$ is usually referred to as the Moss-Burstein (MB) shift~\cite{BursteinPR54}, which is characteristic of all direct-gap degenerate semiconductors and which allows us to estimate the Fermi energy
in our sample. Assuming the full particle-hole symmetry, we obtain $E_F$ below 10~meV. This result is consistent with the Fermi energies found in quantum oscillations experiments performed on Bi$_2$Te$_3$ crystals with similar hole densities~\cite{Kohlerpssb76II}.

\begin{figure}
      \includegraphics[width=.39\textwidth]{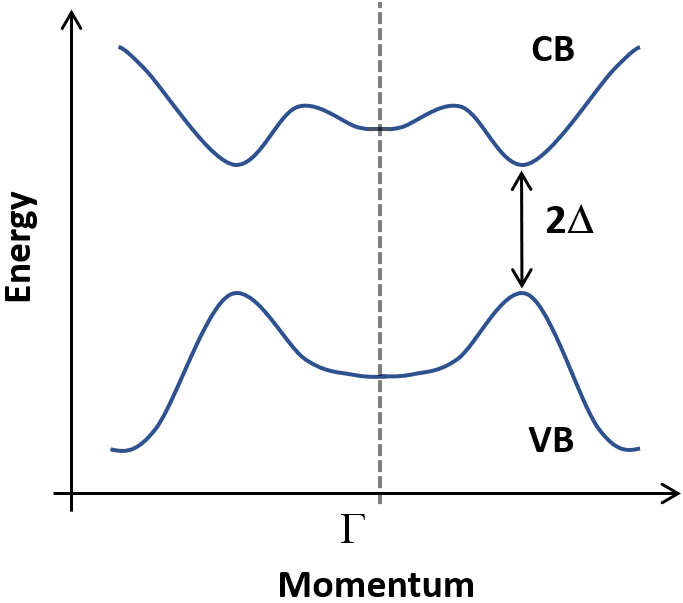}
      \caption{\label{Drawing} Schematic drawing of a profile of the lowest conduction and top-most valence band (CB and VB) with a direct band gap and global extrema located away from $\Gamma$ point that is consistent with our findings, but also with conclusions of magneto-transport experiments and ARPES data. The momentum
      is measured along the line within the mirror planes of Bi$_2$Te$_3$ different from the $c$-axis.}
\end{figure}

Similar to conventional semiconductors, the position of the absorption onset shifts to lower energies with increasing $T$ (inset of Fig.~\ref{Zero-field}).
In this way, we may estimate that, at higher temperatures, the energy band gap
shrinks roughly linearly with $T$: $E_g(\mathrm{meV})\approx 175-0.07\times T [\mathrm{K}]$. Interestingly, the optical conductivity remains non-zero, nearly temperature independent and flat at photon energies below $E_g^*$. This may be assigned to excitations from/to localized states in the band gap, but also, to the absorption tail of the (bulk) free holes whose scattering rate at these frequencies may be limited by disorder.

Above the optical band gap, $\sigma_1(\omega)$ in Bi$_2$Te$_3$ reaches significantly larger values as compared to the sister compound Bi$_2$Se$_3$ (see Fig.~\ref{Zero-field} and Ref.~\cite{DubrokaPRB17}). The latter has an inverted direct band gap with only a slightly higher width, located at the $\Gamma$ point  ($N=1$)~\cite{MartinezSR17}, and the electronic bands are also fairly well described by the Dirac Hamiltonian for massive electrons~\cite{OrlitaPRL15}. The difference in $\sigma_1(\omega)$ thus must lie in specific band structure parameters -- the reduced mass $\mu = m_em_h/(m_e+m_h)$ and/or the valley degeneracy $N$, in particular. Assuming strictly parabolic profiles of bands, the optical conductivity can be approximated using the text-book expression for a direct-gap semiconductor~\cite{YuFS96}:
\begin{equation}
\label{Cardona}
\sigma_1(\omega)~\propto~N \mu^{3/2}\sqrt{\hbar\omega-E_g}.
\end{equation}
Comparing the reduced masses only, $\mu_{\mathrm{Bi_2Se_3}}=0.08m_0$ and $\mu_{\mathrm{Bi_2Te_3}}=0.035m_0$,
one expects greater $\sigma_1$ for Bi$_2$Se$_3$, roughly by a factor of 3. In experimental data, however, an opposite behaviour is observed (Fig.~\ref{Zero-field}). Above the corresponding optical band gaps, $\sigma_1(\omega)$ for Bi$_2$Se$_3$ is roughly by a factor of 5 smaller as compared to Bi$_2$Te$_3$.

This opposite trend, as compared to the one with effective masses, is to a big part due to the large degeneracy $N$, thus confirming the multivalley nature of the band gap concluded in magneto-transport studies~\cite{MallinsonPR68,Kohlerpssb76,Kohlerpssb76II,Kohlerpssb76III}. A simple argumentation based on Eq.~\ref{Cardona} and the overall observed absorption would favorize the degeneracy $N=12$ over 6. However,  we stay rather conservative about this conclusion due to several other possible factors influencing the magnitude of absorption.

Firstly, the impact of the pronounced anisotropy along the $c$-axis ($\mu^{3/2}\rightarrow \mu_\|\mu_c^{1/2}$) was completely neglected. The anisotropy in Bi$_2$Te$_3$ may considerably differ from that in Bi$_2$Se$_3$ and it cannot be deduced from the presented magneto-optical data, which provide us only with the in-plane (i.e., perpendicular to the trigonal axis) estimate of the velocity parameter/effective mass.

Secondly, and likely more importantly, the absorption can be influenced by a non-parabolicity of the bands. Particularly in Bi$_2$Te$_3$, a large non-parabolicity of the conduction band (flattening) was observed 20-30~meV above the band-edge~\cite{Kohlerpssb76,Kohlerpssb76II,Kohlerpssb76III}. This large flattening,
and the corresponding increase of the joint density of states, may enhance the absorption
considerably. This could make the optical data compatible with the $N=6$ valley degeneracy that was concluded, for both the conduction and valence bands, in de Haas-van Alphen and Shubnikov-de Haas studies~\cite{MallinsonPR68,Kohlerpssb76,Kohlerpssb76II, Kohlerpssb76III}.

Combining the this sixfold-degeneracy deduced from quantum oscillations with the direct band gap deduced from magneto-optics, we may schematically sketch the profile of the conduction and valence bands
as a function of the momentum along a line belonging to a mirror plane, but different from the trigonal axis (Fig.~\ref{Drawing}). This drawing respects the possible presence of a local minimum of the bulk conduction band at the $\Gamma$ point indicated by some ARPES studies.

The applicability of the simple massive Dirac model to the magneto-optical response of Bi$_2$Te$_3$ may be somewhat surprising. In fact, the generic two-band models for 2D or 3D topological insulators \cite{BernevigScience06,KonigScience07,ZhangNaturePhys09,LiuPRB10}
always comprise quadratic dispersive diagonal elements, $\Delta \rightarrow \Delta + Mk^2$ which account for the band inversion ($\Delta\cdot M<0$). Such dispersive diagonal elements are responsible for the appearance of the surface states, but they also profoundly impact bulk properties. For instance, when the magnetic field is applied, they lead to characteristic (anti)crossing of zero-mode ($n=0$) LLs in all inverted systems~\cite{KonigScience07,OrlitaPRB11,OrlitaPRL15,AssafPRL17,KrizmanPRB18}.

Even though Bi$_2$Te$_3$ is a topological insulator -- with
experimentally confirmed surface states~\cite{HsiehNature09,ChenScience09,HsiehPRL09} -- such diagonal dispersive terms are not included in our model, which thus keeps the simplest possible massive-Dirac form. This is, for instance, seen from the deduced electron and hole masses that approach very well the Dirac mass $m_D$. This contrasts with Bi$_2$Se$_3$, where the massive Dirac picture is also valid~\cite{OrlitaPRL15}, nevertheless, the presence of the dispersive diagonal elements in the Hamiltonian enhances the mass of electrons and holes by a factor of two as compared to the Dirac mass.

Let us emphasize that the absence of the dispersive terms on the diagonal of the Hamiltonian \eqref{H0} is in line with findings of theoretical studies (see, e.g.,~\cite{AguileraPRB13,AguileraPRB19}). These indicate that the band inversion is present only in a relatively narrow momentum range around the center of the Brillouin zone. The locations of the fundamental band gap thus does not have to coincide with the region of the band inversion, which is located at $\Gamma$ point and which gives rise to the conical band on the surface (consequently centered at the $\overline{\Gamma}$ point of the surface Brillouin zone).

Another implication of the non-inverted direct band gap in Bi$_2$Te$_3$ is that the velocity parameter $v_D$ deduced from our experiments is not the one which determines the slope of the Dirac cones on the surface. Again, this is in contrast to other topological insulators such as Bi$_2$Se$_3$ or Bi$_{1-x}$Sb$_x$ where the regions of the fundamental band gap and of the band inversion overlap, and where, the slope of the surface conical band provides us with a good estimate for the velocity parameter in the Hamiltonian describing bulk states.

\section{Conclusions}

We conclude that bismuth telluride is a direct-gap semiconductor with the band gap of {$E_g=(175\pm 5)$~meV} at low temperatures. The performed analysis of the magneto-optical response implies that the fundamental
band gap is not located on the trigonal axis, which implies its multiple degeneracy ($N=6$ or 12). This conclusion corresponds well with findings of quantum oscillations experiments which indicate the valley degeneracy $N=6$ for both electrons and holes, located in the mirror planes~\cite{MallinsonPR68,Kohlerpssb76,Kohlerpssb76II,Kohlerpssb76III}.
We also conclude that the low-energy electronic excitations in Bi$_2$Te$_3$ are fairly well described within the model of massive Dirac electrons, which comprises only two material parameters.

\begin{acknowledgments}
The authors acknowledge helpful discussions with I.~ Aguilera, D.~M.~Basko, M.~Potemski and  J.~Sanchez-Barriga.
This work was financially supported by the European
Regional Development Fund Project CEITEC Nano+
(No. CZ.021.01/0.0/0.0/16\_013/0001728).
CzechNanoLab project LM2018110 funded by MEYS CR is also gratefully acknowledged for the financial support of the measurements at CEITEC Nano Research Infrastructure. This work was supported by the ANR DIRAC3D project (ANR-17-CE30-0023). A.D. acknowledges support by the Czech Science Foundation (GA\v{C}R) under Project No. GA20-10377S and G.S. by the Austrian Science Fund FWF (Project No. I 4493).
The authors also acknowledge the support of LNCMI-CNRS, a member of the European Magnetic Field Laboratory (EMFL).
\end{acknowledgments}

\section*{Appendix}

\begin{figure}[t]
	\hspace*{-0.8cm}
	\includegraphics[width=.56\textwidth]{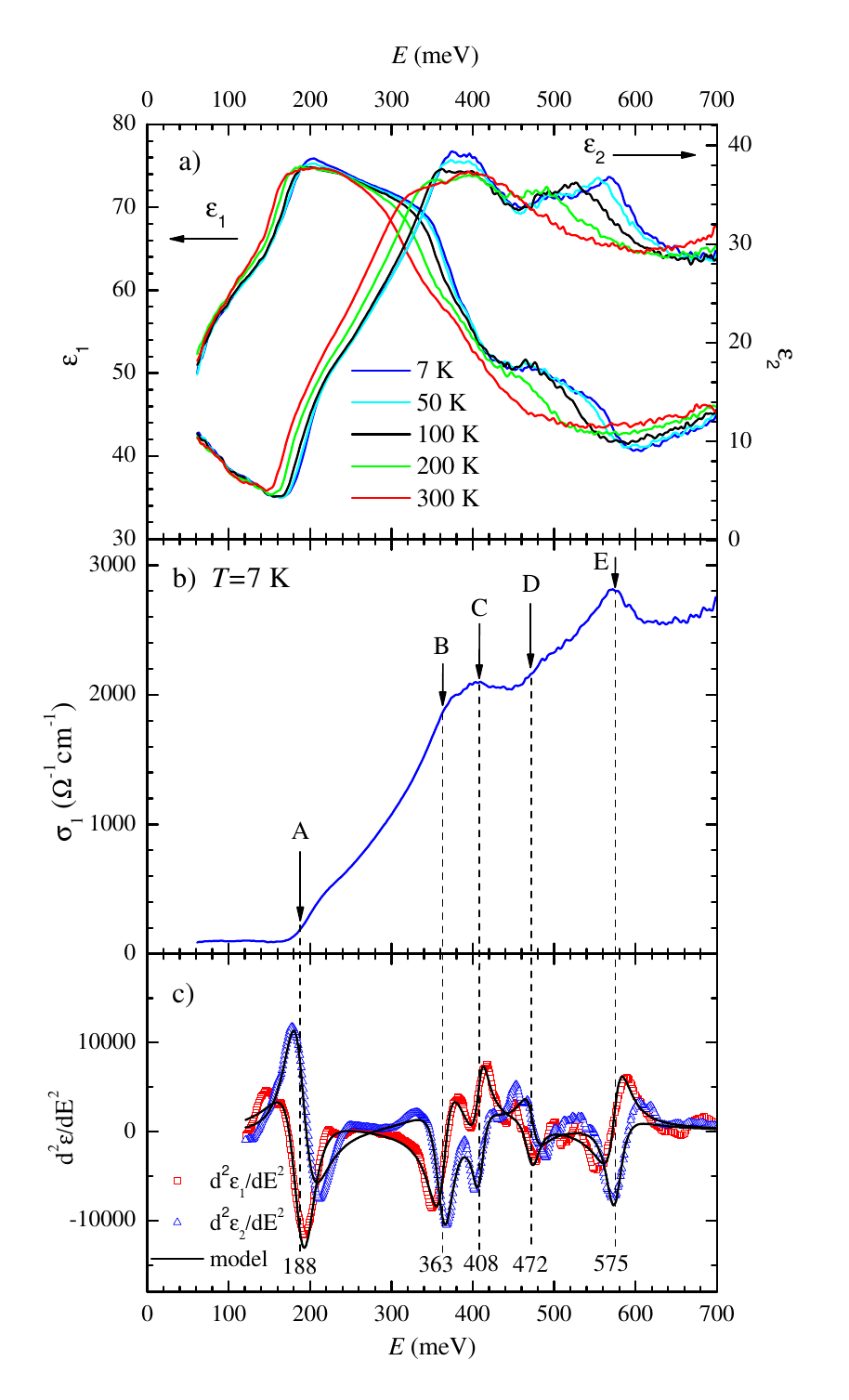}%
	\caption{\label{optics} Zero-field optical data obtained by ellipsometry. Panel (a): the real and imaginary part of the dielectric function. Panel (b): the real part of the optical conductivity at 7~K. The arrows denotes the critical points. Panel (c): displays the second derivative of the real (red squares) and imaginary  (blue triangles) part of the dielectric function and the model spectrum (black line). The numbers are the energies (in meV) of the related critical points.}
\end{figure}

To complement the magneto-optical experiments, we have also performed ellipsometric measurements in the infrared spectral range and analyzed the obtained data using the approach adopted in our preceding
study~\cite{DubrokaPRB17}. The modeling of the optical response enabled us to determine the thickness of the film ($300$~nm)  and the thickness of the surface roughness effective layer. As a second step, the point by point dielectric function was obtained. The deduced real and imaginary part of the dielectric function $\varepsilon=\varepsilon_1+{\rm i}\varepsilon_2$ is shown in Fig.~\ref{optics}a. The related real part of the low-temperature optical conductivity $\sigma_1=\varepsilon_0\omega
\varepsilon_2$ is shown in Fig.~\ref{optics}b.

We have analyzed the interband transitions using the critical point (CP) model~\cite{Cardona1969,LautenschlagerPRB87,Humlicek2002} that is widely applied to the second (or third) derivative of the dielectric function in order to enhance the CPs with respect to the background. The contribution of a parabolic CP to the second derivative is modelled as
\begin{equation}
\label{CP}
\frac{{\rm d}^2\varepsilon}{{\rm d}E^2}=
Ae^{{\rm i}\phi}(E-E_{\rm CP}+{\rm i}\zeta)^{n-2} \;,
\end{equation}
where $A$ is the amplitude, $E_{\rm CP}$ is the energy, $\zeta$ is the broadening and $\phi$ is the phase factor. The exponent $n$ has the values 1/2, 0, -1/2 for three-, two- and one-dimensional CP, respectively. In the simplest case of uncorrelated one-electron bands, the phase $\phi$ takes values of the integer multiples of $\pi/2$. For $A>0$ and 3D critical point ($n=1/2$), the phases $\phi=0$, 90, 180 and 270~deg correspond to $M_1$, $M_2$, $M_3$ and $M_0$ critical points, respectively.  For $A>0$ and a 2D critical point ($n=0$), $\phi=0$, 90 and 180~deg correspond to a minimum ($M_0$), saddle point  ($M_1$) and maximum ($M_2$), respectively~\cite{LautenschlagerPRB87}. However, note that the phase can depart from these integer values for various reasons, e.g., when excitonic effects take place~\cite{RowePRL70}.

\begin{table}[b]
	\caption{\label{Table}%
		The values of the amplitude $A$, energy $E_{\rm CP}$, broadening $\zeta$ and phase $\phi$ obtained from the fit of the CP model to the data shown in Fig.~\ref{optics}c}
	\begin{ruledtabular}
		\begin{tabular}{cccccc}
		label &	$A$ &     $E_{\rm CP}$ & $\zeta$   & $\phi$  & Line shape\\
		&	&    [meV]     & [meV] & [deg]    &  \\
			\colrule
		A&	7.8  &    188&    24&    -29& 2D  \\
		B&	21 eV$^{-1/2}$ &         363&    16&    23&    3D \\
		C&	8 eV$^{-1/2}$ &        408&    11&    76& 3D  \\
		D&	6 eV$^{-1/2}$ &    472&    13&    300& 3D  \\
		E&	16 eV$^{-1/2}$ &      575&    15&    60&    3D \\
		\end{tabular}
	\end{ruledtabular}
\end{table}

The model spectrum fitted to the second derivative of the real and imaginary part of the dielectric function is displayed in Fig.~\ref{optics}c and the obtained values of parameters are shown in Table~\ref{Table}. The main features of the second derivative are the same as the one reported in Ref.~\cite{DubrokaPRB17}, however, here the higher quality of the sample, and consequently of the data, enabled us to resolve five critical points.

Concerning the critical point A at 188~meV, the best fit to the data was obtained using a 2D CP profile that is expected for the  absorption edge influenced by the MB effect, see, e.g., the case of the Bi$_2$Se$_3$ thin film in Ref.~\cite{DubrokaPRB17}. Alternatively, this CP can be modelled with a 3D critical point (albeit with a minor increase in the mean square error) yielding $A=30$~eV$^{-1/2}$, $E_{\rm CP}=188$~meV, $\zeta=17$~meV and the phase $\phi=285$~deg with a relatively minor departure from 270~deg of a simple $M_0$ CP. Obviously, for a small MB shift in the same range as broadening $\zeta$, a mixture of both types of critical points can be expected, which we believe, is the case of the present data. Regardless of the dimensionality, the center energy of the lowest critical point is 188~meV, which is somewhat smaller than 202~meV reported on a similar sample in Ref.~\cite{DubrokaPRB17}. This is presumably caused by a lower doping and correspondingly smaller MB shift. Indeed, the square plasma frequency 4.4$\times10^6$~cm$^{-2}$ (as determined from the
far-infrared reflectivity at 300~K, not shown) is smaller compared to the value of 5.4$\times10^6$~cm$^{-2}$ obtained on the sample studied in Ref.~\cite{DubrokaPRB17}.
The values of $\zeta$ in the range of tens of meV are likely related to sample inhomogeneities, e.g., fluctuations of the local hole density and related fluctuations of the magnitude of the MB effect.

\begin{figure}[t]
	\includegraphics[width=.44\textwidth]{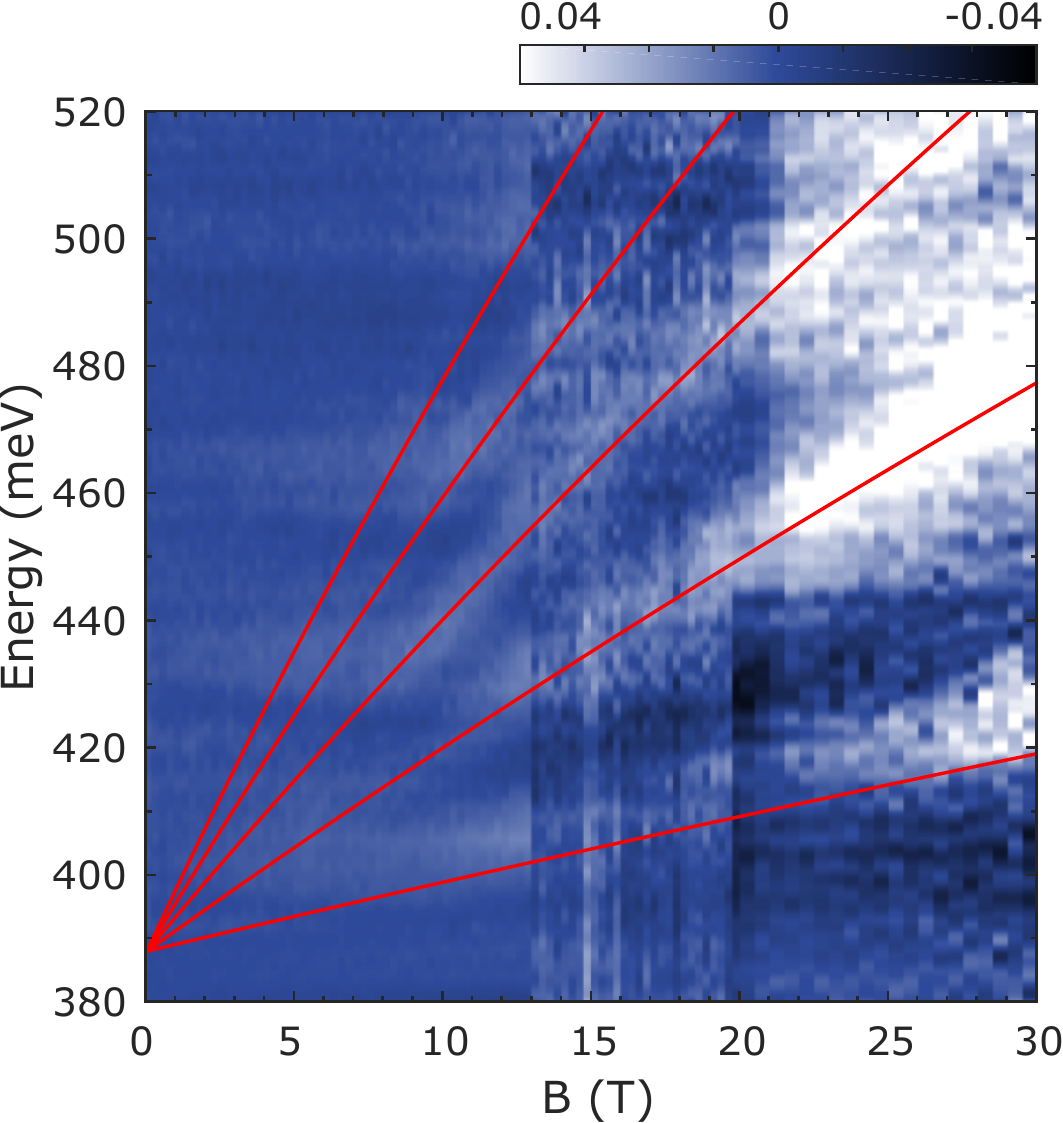}
	\caption{\label{High-energy} False-color plot of relative
		magneto-absorbance in the spectral interval well above the $E_g$. The observed series of inter-LL resonance promote electrons from/to a lower/higher lying band which is not included in the two-band model presented in the main text.}
\end{figure}

The CPs at higher energies were modelled with the 3D CP profiles. The phase values of CPs labelled as B and C (see Tab.~\ref{Table}) suggest that they correspond to the $M_1$ and $M_2$ CPs, respectively, as expected  following the $M_0$ CP. The phase of the critical point D is 300~deg, which is close to 270~deg and which suggests that it corresponds to $M_0$ CP. The phase of the critical point E is 60~deg, which is closest to the $M_2$ critical point. However, it is not excluded that the $M_1$ critical point is somewhat below this energy which effectively reduces the phase from 90~deg.

The presence  of another CP is manifested in the magneto-optical data by appearance of a set of additional weak inter-LL transitions at higher energies (see Fig.~\ref{High-energy}). These  correspond to excitations to/from a more distant band which is not included in the simplified model in the main text. The positions of these additional lines can be reproduced using an ad-hoc invoked massive Dirac Hamiltonian with the gap of 388~meV and the velocity parameter of $6\times10^5$~m/s. This additional CP thus corresponds to another onset of interband excitations, and therefore, it should be of the 3D-$M_0$ type. Nevertheless, it is too weak to be identified directly in the zero-field optical response.

An intriguing question is whether one of experimentally identified CPs arises due to excitations at the $\Gamma$ point. Inspecting the ARPES data~\cite{MichiardiPRB14}, the onset of bulk interband transitions from the upward curved valence band to the conduction band is expected around the energy of 450~meV. Depending on the profile of the conduction and valence bands, the onset may give rise to a 3D critical point with a $M_0$ or $M_3$ character. In the given range of energies,
we indeed find the critical point D  which is of the $M_0$ type. Nevertheless, its
assignment to the onset of interband excitations at the $\Gamma$ point cannot
be more than tentative at the moment.
The ARPES data also suggest that there might be additional critical point(s) in the energy range of 300-400~meV, associated with transitions along the $Z-\Gamma-Z$ axis, in line with our optical data.


%

\end{document}